%% file: conf.tex
\documentclass[preprint,prd,tightenlines,superscriptaddress]{revtex4-1}

\usepackage{graphicx} 
\usepackage{dcolumn}  
\usepackage{colordvi}
\usepackage{color}
\usepackage{epstopdf}
\usepackage{amssymb}
\usepackage{url}
\graphicspath{{ps}}
\usepackage{hyperref}
\usepackage{tabularx}

\newcommand{\benu}{\ensuremath{\Bzb \to D^{*+} e^{-} \nub_e}\xspace}
\newcommand{\bmunu}{\ensuremath{\Bzb \to D^{*+} \mu^{-} \nub_\mu}\xspace}
\newcommand{\btaunu}{\ensuremath{\Bzb \to D^{(*)+} \tau^{-} \nub_\tau}\xspace}
\newcommand{\blnu}{\ensuremath{\Bzb \to D^{*+} \ell^{-} \nub_l}\xspace}

\newcommand{\ifb}{\ensuremath{{\rm fb}^{-1}}\xspace}
\newcommand{\lumi}{\ensuremath{8.70 \pm 0.09~\ifb}\xspace}
\newcommand{\BR}{{\ensuremath{\cal B}}}
\newcommand{\pis}{{\ensuremath{\pi_s}}\xspace}
\newcommand{\costby}{\cos\theta_{BY}}

\input ./belle2sym.tex

\begin{document}

\def\belletwo {\it {Belle II}}

\vspace*{-3\baselineskip}
\resizebox{!}{3cm}{\includegraphics{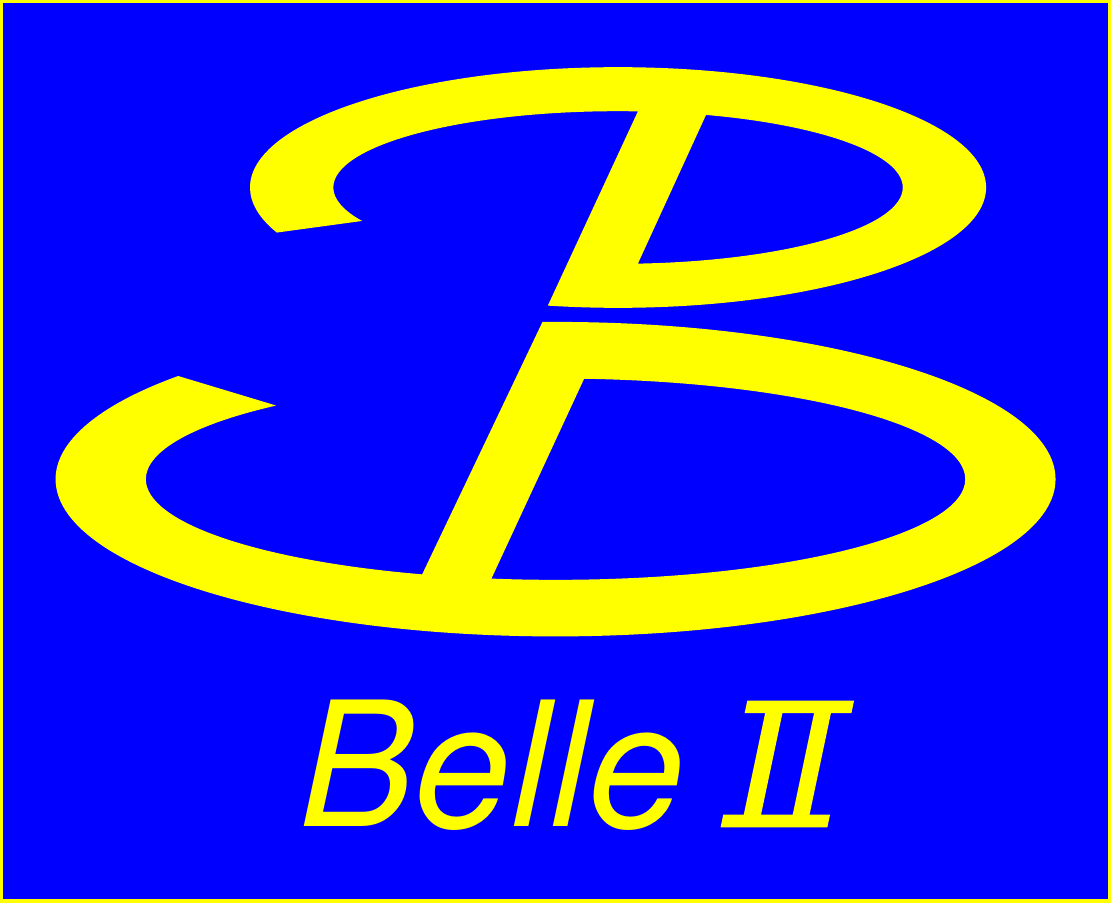}}

\vspace*{-5\baselineskip}
\begin{flushright}
BELLE2-CONF-PH-2020-002 \\
\end{flushright}

\title { \quad\\[0.5cm] Measurement of the branching fraction ${\cal B}(\bar B^0\to D^{*+}\ell^-\bar\nu_\ell)$ with early Belle~II data}

\input{authors-conf2020}

\begin{abstract}
We measure the branching fractions for the decays \benu and \bmunu using \lumi of data collected by the Belle~II experiment at the SuperKEKB asymmetric-energy $e^+ e^-$ collider. Candidate signal decays are reconstructed with the subsequent decays $D^{*+}\to D^0 \pi^+$ and $D^0\to K^-\pi^+$.
We obtain the results
$\BR(\benu)$ = $(4.42\pm0.14(\mathrm{stat})\pm0.33 (\mathrm{syst})) \%$ and $\BR(\bmunu)$ = $(4.70\pm 0.13(\mathrm{stat})\pm0.35(\mathrm{syst})) \%$, 
in agreement with the world averages.
The measurements serve to validate the full chain of detector operation and calibration, data collection and processing, and production of physics results in the case of semileptonic $B$-meson decays.

\end{abstract}


\maketitle

{\renewcommand{\thefootnote}{\fnsymbol{footnote}}}
\setcounter{footnote}{0}


\section{Introduction}
The decays \blnu (where $\ell$ = $e$ or $\mu$) are important for measuring the magnitude of the Cabibbo-Kobayashi-Maskawa matrix element $|V_{cb}|$ and for understanding the hadronic dynamics of $B$ decays. They also constitute a source of background for charmless semileptonic decays and for the study of \btaunu. All this motivates precise measurements of their branching fractions and kinematic distributions. 

The most precise measurements of $\BR(\blnu)$ were obtained by the \babar~\cite{Aubert:2007rs} and Belle~\cite{Waheed:2018djm} collaborations with data sets recorded a decade ago.
Since March 2019, the Belle~II experiment has been collecting $e^+ e^-$ collision data with the full detector. The integrated luminosity is still much smaller than those of the aforementioned experiments. However, this measurement of $\BR(\blnu)$ serves to validate the full analysis chain, from detector operation and calibration, through data collection and processing, to production of physics results, including systematic uncertainties.

\section{The Belle~II detector and data sample}
The Belle~II detector~\cite{Abe:2010sj, ref:b2tip} operates at the SuperKEKB asymmetric-energy  electron-positron collider~\cite{superkekb}, located at the KEK laboratory in Tsukuba, Japan. 
The detector consists of several nested detector subsystems arranged around the beam pipe in a cylindrical geometry. 

The innermost subsystem is the vertex detector, 
which includes two layers of silicon pixel detectors and four outer layers of silicon strip detectors.
Currently, the second pixel layer is installed in only a small part of the solid angle, while the remaining vertex detector 
layers are fully installed. 
Most of the tracking volume consists of a helium and ethane-based small-cell drift chamber. 

Outside the drift chamber, 
a Cherenkov-light imaging and time-of-propagation detector 
provides charged-particle identification in the barrel region. In the forward endcap, this function is provided by a proximity-focusing, ring-imaging Cherenkov detector with an aerogel radiator. 
Further out is an electromagnetic calorimeter, 
consisting of a barrel and two endcap sections made of CsI(Tl) crystals. A uniform 1.5~T magnetic field is provided by a superconducting solenoid situated outside the calorimeter. 
Multiple layers of scintillators and resistive plate chambers, located between the magnetic flux-return iron plates, constitute the $K_L$ and muon 
identification system.


The data used in this analysis were collected at a center-of-mass (CM) energy of 10.58~GeV, corresponding to the mass of the $\Upsilon$(4S) resonance. 
The energies of the electron and positron beams are $7\gev$ and $4\gev$, respectively, resulting in a boost of $\beta\gamma = 0.28$ of the CM frame relative to the lab frame. 
The integrated luminosity of the data is \lumi.

Simulated Monte Carlo (MC) samples of signal events containing $10^5$ events for each of \benu and \bmunu, with the subsequent decays $D^{*+}\to D^0 (\to K^- \pi^+) \pi^+$, are used to obtain the reconstruction efficiencies and signal kinematic distributions. These events were generated with EvtGen~\cite{Lange:2001uf}. Samples of background events are used to obtain kinematic distributions of the background. These include a sample of $e^+ e^-\to B\bar B$ with generic $B$-meson decays, generated with EvtGen, and corresponding to an integrated luminosity of 100~\ifb. A 100~\ifb sample of continuum $e^+e^-\to q\bar q~ (q = u, d, s, c)$ is simulated with KKMC~\cite{Ward:2002qq} interfaced with PYTHIA~\cite{Sjostrand:2007gs}.
All data were analyzed (and, for MC events, generated and simulated) in the basf2~\cite{basf2} framework.

\section{Event selection}

We reconstruct candidate \blnu decays, with the subsequent decays $D^{*+}\to D^0 \pi^+_s$, $D^0\to K^- \pi^+$. Here, $\pis$ indicates the soft pion originating from the $D^{*+}$ decay. Reconstruction of the charge-conjugate decays is implied.
 
Signal candidate reconstruction begins with the selection of four charged-particle tracks. 
The distance of closest approach between each track and the interaction point is required to be less than 2 cm along the $z$ direction (parallel to the beams) and less than 0.5 cm in the transverse $r-\phi$ plane.
%
All tracks except the \pis candidate must have at least one hit in the drift chamber.
The lepton candidate must have a CM frame momentum in the range $p_\ell^* \in [1.2,2.4]$~\gevc. It must also
satisfy lepton-identification (lepton-ID) criteria based on information from all available detectors.
Given the high purity of the decay chain, application of kaon or pion identification criteria is deemed unnecessary and is thus not performed. 

A vertex fit is applied to the $D^0$ candidate, constraining its $K^-\pi^+$ daughter tracks to originate from a common point. The invariant mass of the $D^0$ candidate is required to satisfy \mbox{$m_{K\pi} \in [1.85, 1.88]$~\gevcc} after the fit. The $D^{*+}\to D^0\pi^+_s$  candidate decay is also subjected to a vertex fit, after which the mass difference between the $D^*$ and $D^0$ candidates is required to satisfy $\Delta m \in [0.144,0.148]$~\gevcc. 

Continuum background is suppressed by requiring the momentum of the $D^*$ candidate in the CM frame to be less than 2.5~\gevc. Further continuum suppression is achieved by requiring $R_2 < 0.3$, where $R_2$ is the ratio of the second and zeroth Fox-Wolfram moments~\cite{Fox:1978vu}, calculated using all the tracks and photon candidates in the event.

After applying all the selection criteria above, multiple \blnu candidates are found in about 1\% of the events. In these events, we perform a vertex fit for the decay \blnu and select the candidate with the smallest value of the vertex-fit $\chi^2$. 
The signal efficiency after all selection criteria is 
$(19.8 \pm 1.8)\%$ for \benu and $(21.4 \pm 2.2)\%$ for \bmunu. These values are obtained from signal MC with lepton-ID efficiency corrections obtained from data-MC comparisons of reconstructed $J/\psi\to\ell^+\ell^-$ decays. The quoted uncertainties are dominated by the uncertainties on this correction, and also include the much smaller MC-statistical uncertainty.

\section{Signal yield extraction}

For each candidate, we calculate the quantity
\begin{equation}
\costby \equiv {2E_B^* E_Y^* - M_B^2 - m_Y^2 \over 2 |p_B^*||p_Y^*|},
\end{equation}
where $E_Y^*$, $|p_Y^*|$, and $m_Y$ are the CM energy, momentum, and invariant mass of the $D^{*}\ell$ system, $M_B$ is the nominal $B$ mass~\cite{Zyla:2020},
and $E_B^*$, $|p_B^*|$ are the CM energy and momentum of the $B$; the CM is inferred from the beam four-momenta.  For correctly reconstructed \blnu candidates with perfect detector resolution and correct values of $E_B^*$ and $p_B^*$, $\costby$ is the cosine of the angle between the momentum of the $B$ meson and that of the $D^{*}\ell$ system in the CM frame. Given the finite beam-energy spread, final-state radiation, and detector resolution, the $\costby$ distribution  of signal events extends beyond the range $[-1,1]$. 
The $\costby$ distributions of data and MC events are shown in Fig.~\ref{fig:results}.

We obtain the event yields 
$N_s$, $N_B$, and $N_c$
for signal, $B\bar B$ background, and continuum, respectively, from a binned, extended-maximum-likelihood fit to the $\costby$ distribution of the data. The probability density function (PDF) used in the fit is the sum of the PDFs of these three event categories. These PDFs are obtained from  the MC samples, after application of  momentum- and polar-angle-dependent corrections
to the lepton-identification efficiencies of leptons and hadrons. For leptons, corrections of the order of a few percent are obtained from $J/\psi\to \ell^+\ell^- ~(\ell = e, \mu)$ decays. Corrections for hadrons misidentified as leptons, obtained from samples of reconstructed $D^0\to K^-\pi^+$ decays, are of order unity in some regions. Applying these corrections to the PDFs changes the signal yield from the fit by 0.5\% for \benu and by 0.1\% for \bmunu.

The fit results are shown in Fig.~\ref{fig:results}  and summarized in Table~\ref{tab:results}.

\begin{figure}
\begin{center}
\begin{tabular}{cccc}
\includegraphics[width=0.5\columnwidth]{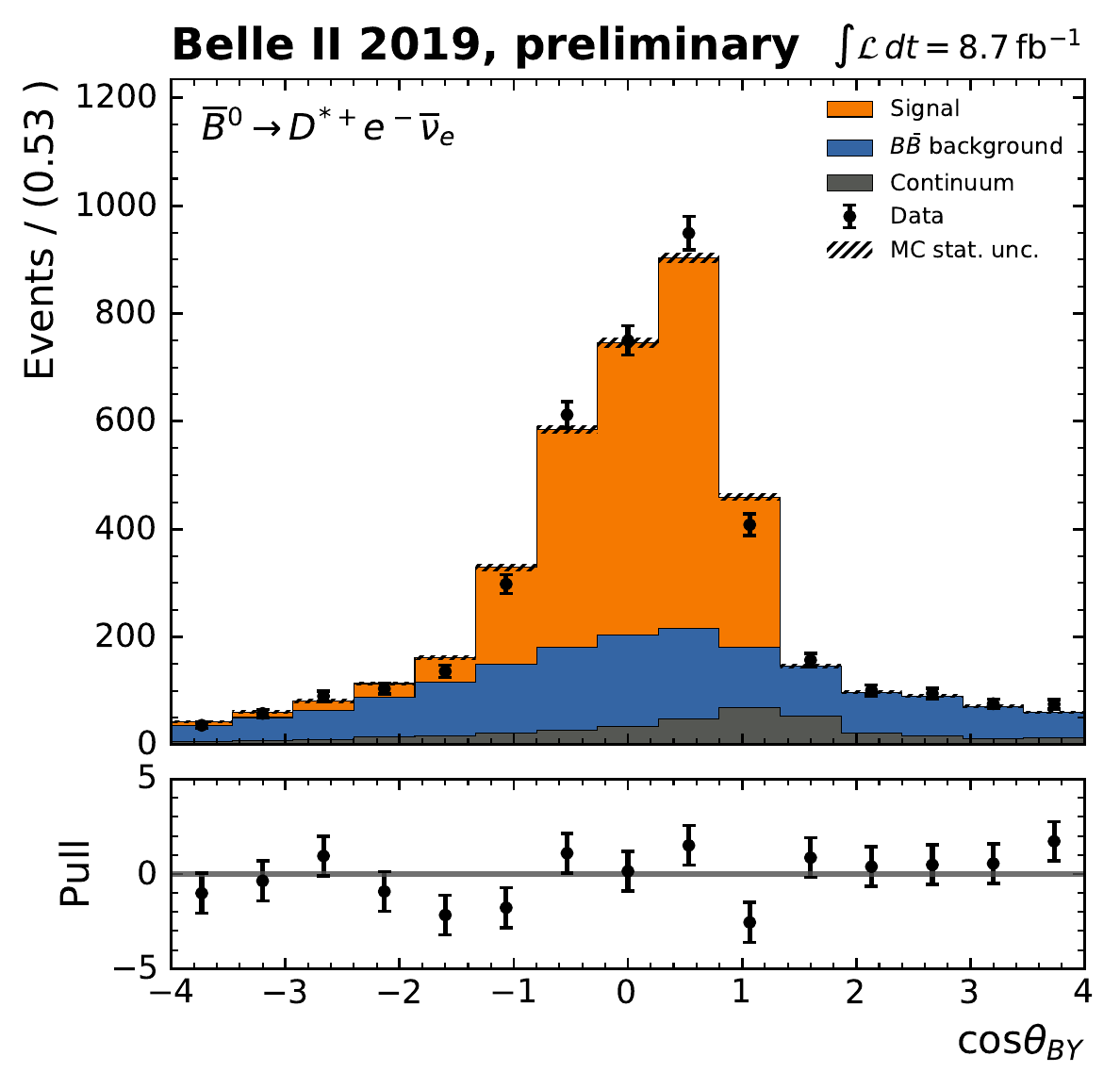} &
\includegraphics[width=0.5\columnwidth]{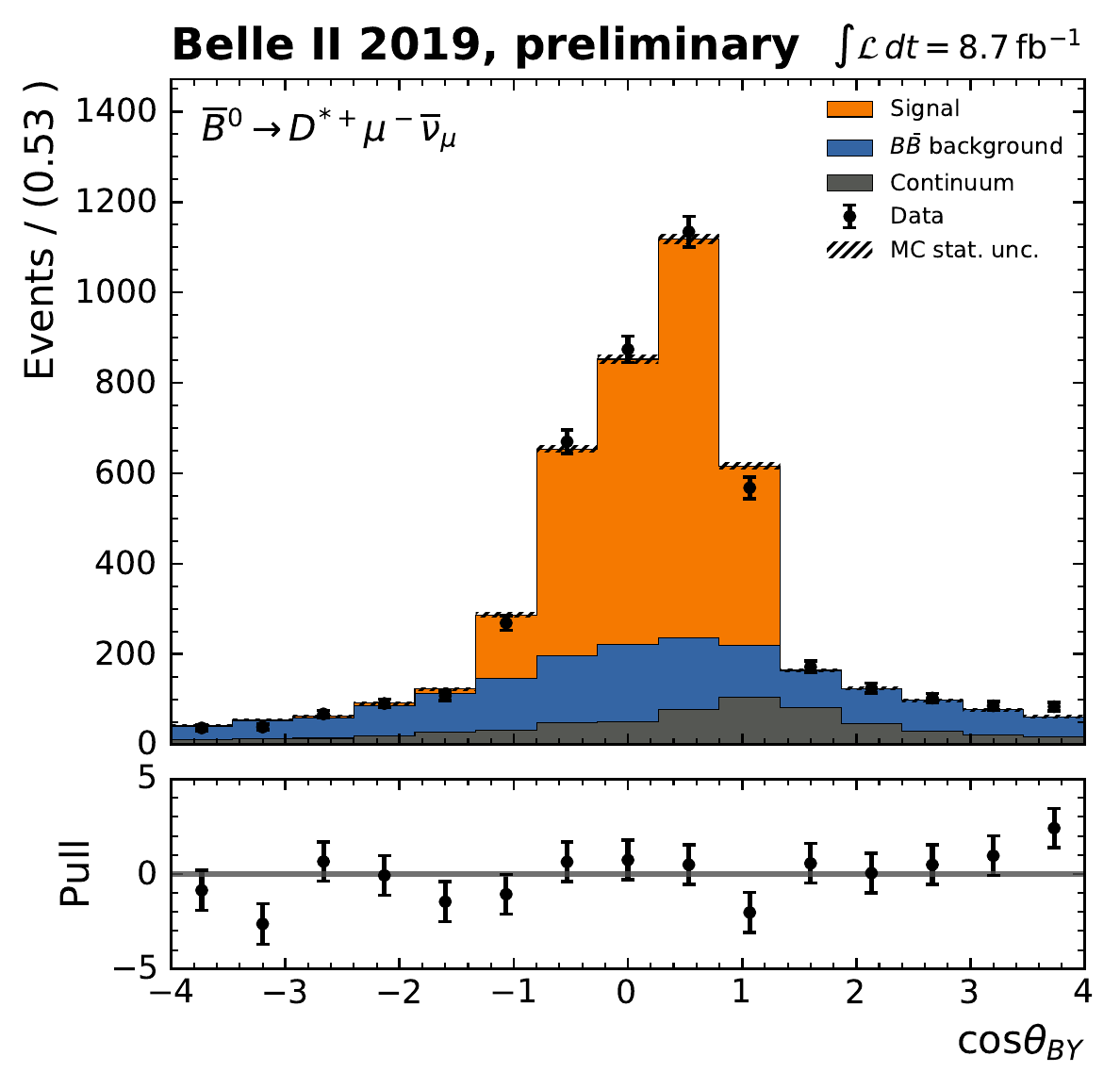}  \\
\end{tabular}
  \caption{Post-fit $\costby$ distributions for the selected \benu (left) and \bmunu (right) candidates (data points), overlaid with the signal, $B\bar B$ background, and continuum PDFs used in the fit as obtained from simulated MC events (histograms). The pull distributions show the difference between the data and total PDF divided by the square root of the number of data events in each bin.
 }
  \label{fig:results}
\end{center}
\end{figure}

\begin{table}[htbp]
\centering
\begin{tabular}{l|c|c}
    \hline\hline
    Event type & \benu Yield & \bmunu Yield \\
    \hline
    $N_s$ (Signal) & $2199\pm 69$  & $2525\pm 68$ \\
    $N_{B^0}$ ($B\bar B$) & ~$1377\pm 123$ &~$1306\pm 129$ \\
    $N_c$ (Continuum) & ~~$369\pm 106$ &~~~$597\pm 116$ \\
    \hline\hline
    \end{tabular}
    \caption{Signal, $B\bar B$, and continuum event yields obtained from the fits shown in Fig.~\ref{fig:results}. The uncertainty is statistical.}
    \label{tab:results}
\end{table}

The branching fraction is obtained from the signal yield via
\begin{equation}
\BR(\blnu)= {N_s \over {\epsilon \times N_{B^0}} \times \BR(D^{*+}\to D^0\pi^+)\times  \BR(D^0\to K^-\pi^+)} \, ,
\label{eq:BR-eq}
\end{equation}
where $\epsilon$ is the product of the signal reconstruction efficiency and acceptance, and $N_{B^0}$ is the number of $B^0$ mesons in the data sample, further discussed in section~\ref{sec:syst}.

\section {Systematic uncertainties and cross checks}\label{sec:syst}

The relative systematic uncertainties affecting the measurement are listed in Table~\ref{tab:syst}. We assume no correlation among the individual sources of uncertainty and sum them in quadrature to obtain the total systematic uncertainty. The  methods used for obtaining these uncertainties are described below.

\begin{table}[htbp]
\begin{center}
\begin{tabular}{l|c|c}\hline\hline
Source & \multicolumn{2}{c}{Relative uncertainty (\%)} \\
    \cline{2-3}
       &\benu & \bmunu \\
\hline
Lepton-ID, PDF & 0.09 & 0.08 \\
MC statistics, PDF  & 0.64 & 0.55 \\
$\BR(\bar B\to D^{**} \ell\bar\nu)$ & 0.18  & 0.08 \\
Efficiency momentum dependence & 0.1 & 0.1 \\
PDF binning & 0.5 & 0.5 \\
Lepton-ID, efficiency & 1.8 & 2.2 \\
MC statistics, efficiency  & 0.13  & 0.13 \\
Tracking of $K$, $\pi$, $\ell$ & 2.5 & 2.5 \\
Tracking of $\pis$ & 6.0 & 6.0 \\
$N_{B^0}$  & 2.7 & 2.7
\\
Charm branching fractions & 1.1  & 1.1 \\
\hline
Total &7.40 & 7.50\\
 \hline\hline
\end{tabular}
\end{center}
\caption{Summary of the relative systematic uncertainties for the measurements of $\BR(\blnu)$.
The first five uncertainties impact the extracted signal yield, while the others impact the other factors of Eq.~(\ref{eq:BR-eq}).
}\label{tab:syst}
\end{table}

The lepton-identification corrections are measured with statistical uncertainties that arise from the limited size of the control samples, as well as systematic uncertainties. We produce 500 sets of correction values sampled from Gaussian distributions that reflect these uncertainties, accounting for systematic correlations. Each set of corrections is used to produce new MC PDFs that are used to refit the data $\costby$ distribution. The standard deviation of the distribution of the signal yield obtained with the 500 sets is 2 events for both \benu and \bmunu. This value is thus used as the associated systematic uncertainty on the signal yield.
The same procedure is used to calculate the correction on the absolute lepton-identification efficiency affecting the overall signal efficiency.

The impact of the finite sizes of the MC samples on the fit result is evaluated by varying the MC PDFs bin-by-bin according to Poisson distributions, and repeating the fit 500 times. We take the standard deviation of the resulting signal-yield distribution to estimate the systematic uncertainty. 

The semileptonic decays $\bar B\to D^{**} \ell\bar\nu$, where $D^{**}$ indicates an excited charm meson heavier than the $D^*$, have a similar particle content to that of signal decays. As a result, the fit may be biased if the branching fractions of $\bar B\to D^{**} \ell\bar\nu$ are incorrect in the generic MC sample. To estimate the systematic uncertainty, we obtain the $B\bar B$ PDF from the MC after varying the branching fractions for these decays by $\pm 25\%$, which is twice the relative uncertainty on $\BR(\bar B\to D^0 \pi^+ \ell^-\bar\nu)$. The resulting change in the signal yield is taken as the systematic uncertainty.

Further Data-MC differences in the reconstruction efficiency as a function of the momenta $p_\ell$ and $p_\pis$ of the lepton and soft-pion may affect the $\costby$ PDFs and hence the signal yield.
Figure~\ref{fig:momenta} shows the $p_\ell$ and $p_\pis$  distributions in the CM frame for data and MC, with the corresponding distributions of the pull $(N_{\rm data}-N_{\rm MC})/\sqrt{N_{\rm data}}$ in each bin.
The $\chi^2$ per degrees of freedom of the pull distributions is $\chi^2/N_{\rm DOF} =1.6$ for Figs.~\ref{fig:momenta}a,~b,~c, and 3.1 for Fig.~\ref{fig:momenta}d, where $N_{\rm DOF}=9$.  
To test the impact of this relative disagreement on the fit, we reproduce the $\costby$ PDFs after reweighting each MC event so as to obtain $\chi^2/N_{\rm DOF} =1$ in the $p_\pis$ and $p_\ell$ pull distributions. Using these modified PDFs to perform the fit leads to a maximal change in the fit yield of 0.1\%, which is taken as a systematic uncertainty. 

\begin{figure}[t]
\begin{center}
\begin{tabular}{cc}
\includegraphics[width=0.45\columnwidth]{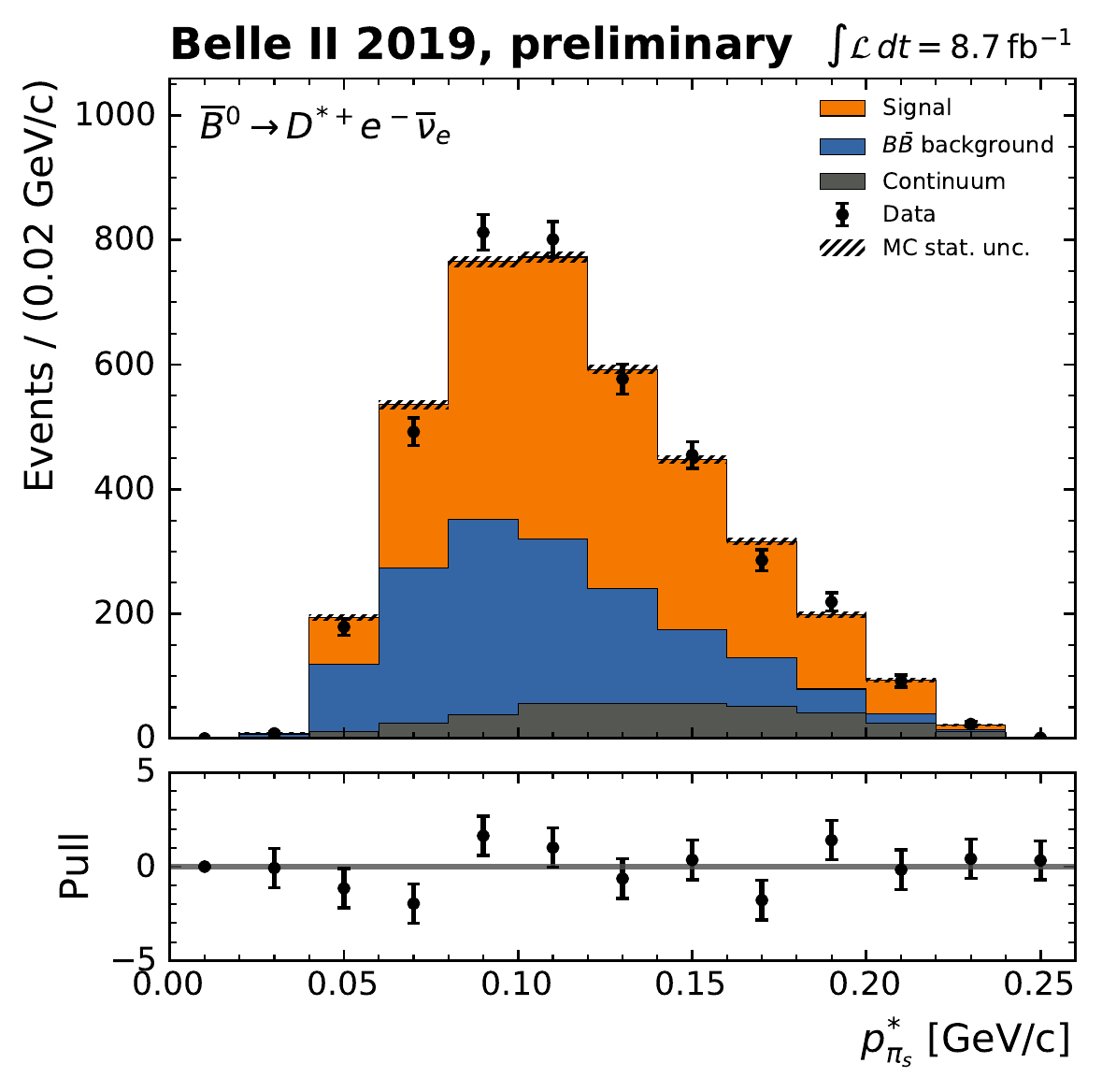} &
\includegraphics[width=0.45\columnwidth]{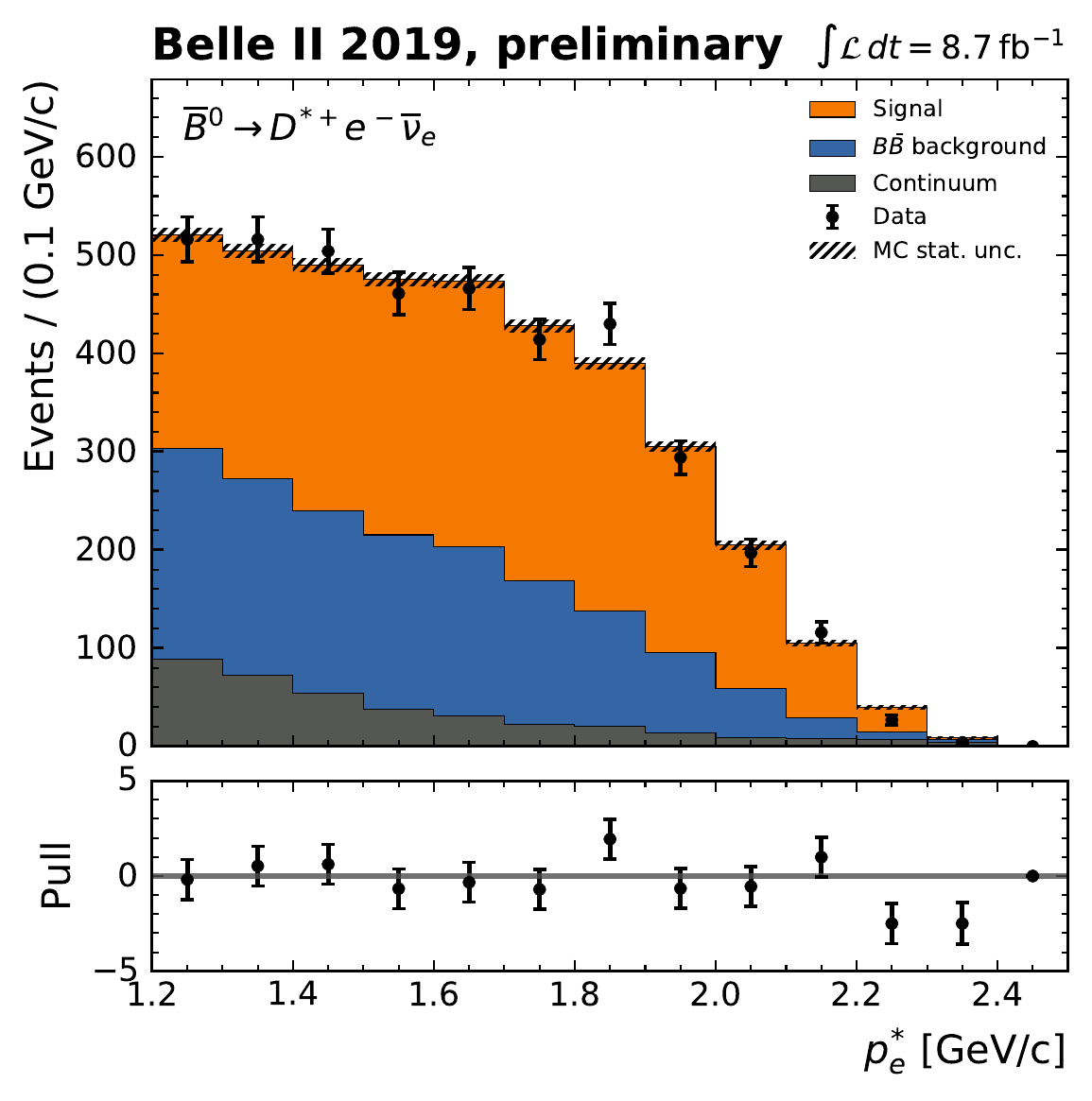} \\
\includegraphics[width=0.45\columnwidth]{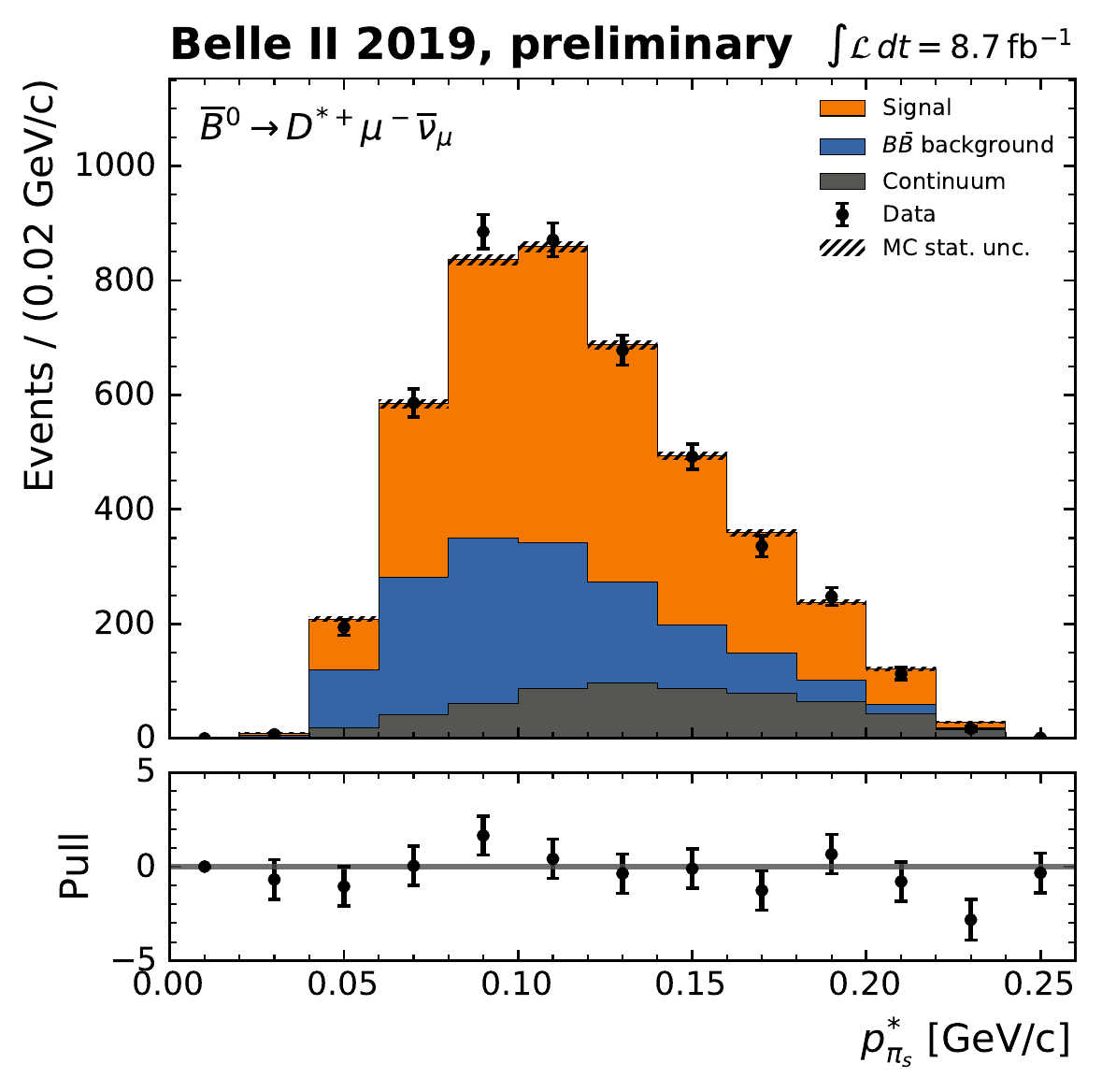} &
\includegraphics[width=0.45\columnwidth]{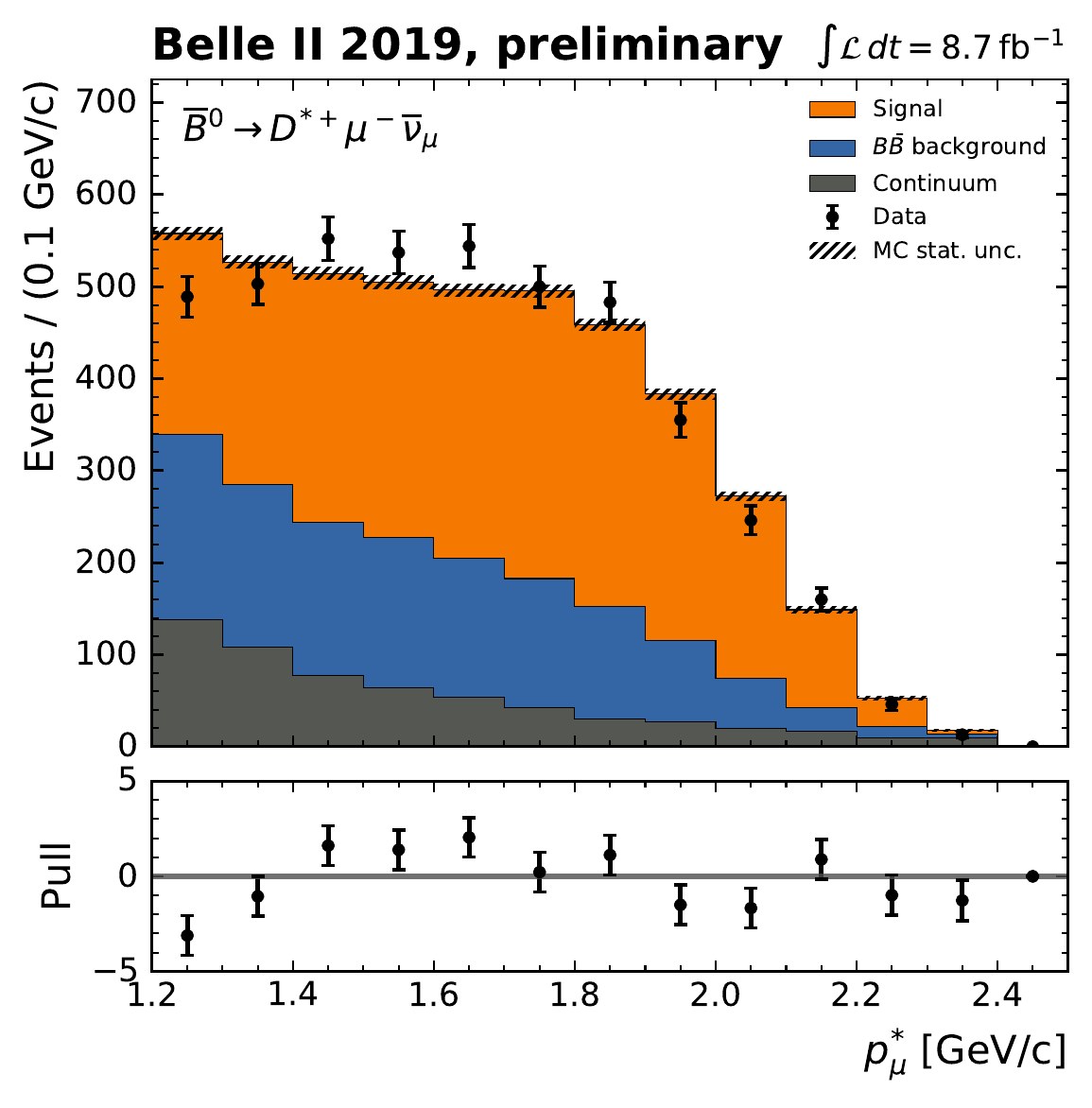} \\
\end{tabular}
   \caption{Distribution of the CM momentum of the slow pion (left plots) and lepton (right plots) for the \benu (top plots) and \bmunu (bottom plots) modes. The data distributions are overlaid with the MC PDFs, scaled according to the fit results.}
 \label{fig:momenta}
\end{center}
\end{figure}

The systematic uncertainty due to the choice of the number of bins in the PDF is estimated to be 0.5\% by varying the number of bins up to 40.

The tracking efficiency uncertainty for the lepton, kaon, and pion is 0.82\% per track. This is obtained by comparing $R_{2/3}$ for $e^+e^-\to \tau^+\tau^-$ events in data and MC, where $R_{2/3}$ is the fraction of 3-prong $\tau$ decays in which only two hadron tracks are found. The uncertainty on the soft pion tracking efficiency, 6.0\%, is calculated as 20\% of the inefficiency obtained from MC. The value of 20\% is taken from data-MC comparison studies performed with high-momentum tracks.

To obtain the number of $B^0$ mesons in the sample, we use the relation 
\begin{equation}
N_{B^0} =  \mathcal{L} \times \sigma(e^+e^-\to \Upsilon(4S))\times 2 \times f^{00} \, . 
\end{equation}
Here $\mathcal{L}$ denotes the integrated luminosity, measured to be \lumi using $e^+e^-\to e^+ e^-$ and $e^+e^-\to \gamma\gamma$ events~\cite{Abudinen:2019osb}. 
The factor $f^{00}=0.487\pm0.013$ is the fraction of neutral $B$ mesons produced in $\Upsilon(4S)$ decays~\cite{Zyla:2020}.
We take the effective cross section  $\sigma(e^+e^-\to \Upsilon(4S))$ to be $1.110\pm 0.008~{\rm nb}$ from measurements of the number of $B$-meson pairs produced~\cite{Bevan:2014iga} and of the integrated luminosity~\cite{Lees:2013rw} at the \babar\ experiment. This assumes that the branching fraction for $\Upsilon(4S)$ decays to non-$B\bar B$ final states is negligible.
We assign an additional uncertainty of 2\% on this value, to account for the possibility that the average SuperKEKB CM energy is off the $\Upsilon(4S)$ resonance by up to 1.5~MeV.

Lastly, we account for the impact of the uncertainties in the charm branching fractions, $\BR(\Dstarp \to \Dz \pip)$ = $(67.7\pm0.5)\%$ and $\BR(\Dz \to \Km \pip)$ = $(3.950\pm0.031)\%$~\cite{Zyla:2020}, on the signal branching fraction.

\section{Branching fraction results and summary}

From Eq.~(\ref{eq:BR-eq}) we measure for the \blnu branching fractions:
\begin{eqnarray}
\BR(\benu)&=& (4.42 \pm 0.14 (\mathrm{stat}) \pm 0.33 (\mathrm{sys}))\% \, ,\\
\BR(\bmunu)&=& (4.70 \pm 0.13 (\mathrm{stat}) \pm 0.35 (\mathrm{sys}))\% \, .
\end{eqnarray}
Considering the statistical uncertainties only, the branching fractions for the two modes are consistent to within 1.5 standard deviations.

In summary, we measure the branching fractions of the decays \benu and \bmunu using 8.7~\ifb of data collected by the Belle~II experiment in 2019. 
The world average value of the branching fraction $\BR(\blnu) =(5.05 \pm 0.14)\%$~\cite{Zyla:2020} was obtained with a different value of the $D^0$ branching fraction, $\BR(D^0\to K^-\pi^+)=(3.89\pm 0.04)\%$.
Accounting for this difference, our results are lower than the world average by 1.5 and 0.7 standard deviation for the \benu and \bmunu modes, respectively. While the uncertainties we estimate are significantly larger than those of the world average, this is the first branching-fraction measurement performed with Belle~II data, and constitutes a test of the entire data production, processing, and analysis chain.

\section{ACKNOWLEDGEMENTS}
\input{acknowledgements}

\bibliography{conf.bib}

\end{document}

%% file: authors-conf2020.tex
\newcommand{\instSinica}{Academia Sinica, Taipei 11529, Taiwan}
\newcommand{\instBeihang}{Beihang University, Beijing 100191, China}
\newcommand{\instBUAP}{Benemerita Universidad Autonoma de Puebla, Puebla 72570, Mexico}
\newcommand{\instBNL}{Brookhaven National Laboratory, Upton, New York 11973, U.S.A.}
\newcommand{\instBINP}{Budker Institute of Nuclear Physics SB RAS, Novosibirsk 630090, Russian Federation}
\newcommand{\instCMU}{Carnegie Mellon University, Pittsburgh, Pennsylvania 15213, U.S.A.}
\newcommand{\instCPPM}{Centre de Physique des Particules de Marseille, 13288 Marseille, France}
\newcommand{\instCinvestavIPN}{Centro de Investigacion y de Estudios Avanzados del Instituto Politecnico Nacional, Mexico City 07360, Mexico}
\newcommand{\instPrague}{Faculty of Mathematics and Physics, Charles University, 121 16 Prague, Czech Republic}
\newcommand{\instChiangMai}{Chiang Mai University, Chiang Mai 50202, Thailand}
\newcommand{\instChiba}{Chiba University, Chiba 263-8522, Japan}
\newcommand{\instChonnam}{Chonnam National University, Gwangju 61186, South Korea}
\newcommand{\instConacyt}{Consejo Nacional de Ciencia y Tecnolog\'{\i}a, Mexico City 03940, Mexico}
\newcommand{\instDESY}{Deutsches Elektronen--Synchrotron, 22607 Hamburg, Germany}
\newcommand{\instDuke}{Duke University, Durham, North Carolina 27708, U.S.A.}
\newcommand{\instDuyTan}{Institute of Theoretical and Applied Research (ITAR), Duy Tan University, Hanoi 100000, Vietnam}
\newcommand{\instENEA}{ENEA Casaccia, I-00123 Roma, Italy}
\newcommand{\instEri}{Earthquake Research Institute, University of Tokyo, Tokyo 113-0032, Japan}
\newcommand{\instJuelich}{Forschungszentrum J\"{u}lich, 52425 J\"{u}lich, Germany}
\newcommand{\instFuJen}{Department of Physics, Fu Jen Catholic University, Taipei 24205, Taiwan}
\newcommand{\instFudan}{Key Laboratory of Nuclear Physics and Ion-beam Application (MOE) and Institute of Modern Physics, Fudan University, Shanghai 200443, China}
\newcommand{\instGoettingen}{II. Physikalisches Institut, Georg-August-Universit\"{a}t G\"{o}ttingen, 37073 G\"{o}ttingen, Germany}
\newcommand{\instGifu}{Gifu University, Gifu 501-1193, Japan}
\newcommand{\instSOKENDAI}{The Graduate University for Advanced Studies (SOKENDAI), Hayama 240-0193, Japan}
\newcommand{\instGyeongsang}{Gyeongsang National University, Jinju 52828, South Korea}
\newcommand{\instHanyang}{Department of Physics and Institute of Natural Sciences, Hanyang University, Seoul 04763, South Korea}
\newcommand{\instKEK}{High Energy Accelerator Research Organization (KEK), Tsukuba 305-0801, Japan}
\newcommand{\instJPARC}{J-PARC Branch, KEK Theory Center, High Energy Accelerator Research Organization (KEK), Tsukuba 305-0801, Japan}
\newcommand{\instHSE}{Higher School of Economics (HSE), Moscow 101000, Russian Federation}
\newcommand{\instIISER}{Indian Institute of Science Education and Research Mohali, SAS Nagar, 140306, India}
\newcommand{\instIITBhubaneswar}{Indian Institute of Technology Bhubaneswar, Satya Nagar 751007, India}
\newcommand{\instIITGuwahati}{Indian Institute of Technology Guwahati, Assam 781039, India}
\newcommand{\instIITHyderabad}{Indian Institute of Technology Hyderabad, Telangana 502285, India}
\newcommand{\instIITMadras}{Indian Institute of Technology Madras, Chennai 600036, India}
\newcommand{\instIndiana}{Indiana University, Bloomington, Indiana 47408, U.S.A.}
\newcommand{\instIHEPRussia}{Institute for High Energy Physics, Protvino 142281, Russian Federation}
\newcommand{\instHEPHYVienna}{Institute of High Energy Physics, Vienna 1050, Austria}
\newcommand{\instIHEPChina}{Institute of High Energy Physics, Chinese Academy of Sciences, Beijing 100049, China}
\newcommand{\instChennai}{Institute of Mathematical Sciences, Chennai 600113, India}
\newcommand{\instIPP}{Institute of Particle Physics (Canada), Victoria, British Columbia V8W 2Y2, Canada}
\newcommand{\instIOP}{Institute of Physics, Vietnam Academy of Science and Technology (VAST), Hanoi, Vietnam}
\newcommand{\instIFIC}{Instituto de Fisica Corpuscular, Paterna 46980, Spain}
\newcommand{\instFrascati}{INFN Laboratori Nazionali di Frascati, I-00044 Frascati, Italy}
\newcommand{\instNapoliINFN}{INFN Sezione di Napoli, I-80126 Napoli, Italy}
\newcommand{\instPadovaINFN}{INFN Sezione di Padova, I-35131 Padova, Italy}
\newcommand{\instPerugiaINFN}{INFN Sezione di Perugia, I-06123 Perugia, Italy}
\newcommand{\instPisaINFN}{INFN Sezione di Pisa, I-56127 Pisa, Italy}
\newcommand{\instRomaINFN}{INFN Sezione di Roma, I-00185 Roma, Italy}
\newcommand{\instRomaTreINFN}{INFN Sezione di Roma Tre, I-00146 Roma, Italy}
\newcommand{\instTorinoINFN}{INFN Sezione di Torino, I-10125 Torino, Italy}
\newcommand{\instTriesteINFN}{INFN Sezione di Trieste, I-34127 Trieste, Italy}
\newcommand{\instJAEA}{Advanced Science Research Center, Japan Atomic Energy Agency, Naka 319-1195, Japan}
\newcommand{\instMainz}{Johannes Gutenberg-Universit\"{a}t Mainz, Institut f\"{u}r Kernphysik, D-55099 Mainz, Germany}
\newcommand{\instGiessen}{Justus-Liebig-Universit\"{a}t Gie\ss{}en, 35392 Gie\ss{}en, Germany}
\newcommand{\instKarlsruhe}{Institut f\"{u}r Experimentelle Teilchenphysik, Karlsruher Institut f\"{u}r Technologie, 76131 Karlsruhe, Germany}
\newcommand{\instKennesaw}{Kennesaw State University, Kennesaw, Georgia 30144, U.S.A.}
\newcommand{\instKitasato}{Kitasato University, Sagamihara 252-0373, Japan}
\newcommand{\instKISTI}{Korea Institute of Science and Technology Information, Daejeon 34141, South Korea}
\newcommand{\instKorea}{Korea University, Seoul 02841, South Korea}
\newcommand{\instKSU}{Kyoto Sangyo University, Kyoto 603-8555, Japan}
\newcommand{\instKyotoU}{Kyoto University, Kyoto 606-8501, Japan}
\newcommand{\instKyungpook}{Kyungpook National University, Daegu 41566, South Korea}
\newcommand{\instLAL}{Laboratoire de l'Acc\'{e}l\'{e}rateur Lin\'{e}aire, IN2P3/CNRS et Universit\'{e} Paris-Sud 11, Centre Scientifique d'Orsay, F-91898 Orsay Cedex, France}
\newcommand{\instLPI}{P.N. Lebedev Physical Institute of the Russian Academy of Sciences, Moscow 119991, Russian Federation}
\newcommand{\instLNNU}{Liaoning Normal University, Dalian 116029, China}
\newcommand{\instLMU}{Ludwig Maximilians University, 80539 Munich, Germany}
\newcommand{\instLuther}{Luther College, Decorah, Iowa 52101, U.S.A.}
\newcommand{\instMNITJaipur}{Malaviya National Institute of Technology Jaipur, Jaipur 302017, India}
\newcommand{\instMPP}{Max-Planck-Institut f\"{u}r Physik, 80805 M\"{u}nchen, Germany}
\newcommand{\instMPGHLL}{Semiconductor Laboratory of the Max Planck Society, 81739 M\"{u}nchen, Germany}
\newcommand{\instMcGill}{McGill University, Montr\'{e}al, Qu\'{e}bec, H3A 2T8, Canada}
\newcommand{\instMETU}{Middle East Technical University, 06531 Ankara, Turkey}
\newcommand{\instMEPhI}{Moscow Physical Engineering Institute, Moscow 115409, Russian Federation}
\newcommand{\instNagoya}{Graduate School of Science, Nagoya University, Nagoya 464-8602, Japan}
\newcommand{\instNagoyaKMI}{Kobayashi-Maskawa Institute, Nagoya University, Nagoya 464-8602, Japan}
\newcommand{\instNaraWu}{Nara Women's University, Nara 630-8506, Japan}
\newcommand{\instUNAM}{National Autonomous University of Mexico, Mexico City, Mexico}
\newcommand{\instNTUTaiwan}{Department of Physics, National Taiwan University, Taipei 10617, Taiwan}
\newcommand{\instNUUTaiwan}{National United University, Miao Li 36003, Taiwan}
\newcommand{\instKrakow}{H. Niewodniczanski Institute of Nuclear Physics, Krakow 31-342, Poland}
\newcommand{\instNiigata}{Niigata University, Niigata 950-2181, Japan}
\newcommand{\instNSU}{Novosibirsk State University, Novosibirsk 630090, Russian Federation}
\newcommand{\instOkinawa}{Okinawa Institute of Science and Technology, Okinawa 904-0495, Japan}
\newcommand{\instOsakaCity}{Osaka City University, Osaka 558-8585, Japan}
\newcommand{\instRCNP}{Research Center for Nuclear Physics, Osaka University, Osaka 567-0047, Japan}
\newcommand{\instPNNL}{Pacific Northwest National Laboratory, Richland, Washington 99352, U.S.A.}
\newcommand{\instPanjab}{Panjab University, Chandigarh 160014, India}
\newcommand{\instPeking}{Peking University, Beijing 100871, China}
\newcommand{\instPanjabPAU}{Punjab Agricultural University, Ludhiana 141004, India}
\newcommand{\instRIKEN}{Theoretical Research Division, Nishina Center, RIKEN, Saitama 351-0198, Japan}
\newcommand{\instXavier}{St. Francis Xavier University, Antigonish, Nova Scotia, B2G 2W5, Canada}
\newcommand{\instSeoul}{Seoul National University, Seoul 08826, South Korea}
\newcommand{\instShandong}{Shandong University, Jinan 250100, China}
\newcommand{\instSPU}{Showa Pharmaceutical University, Tokyo 194-8543, Japan}
\newcommand{\instSoochow}{Soochow University, Suzhou 215006, China}
\newcommand{\instSoongsil}{Soongsil University, Seoul 06978, South Korea}
\newcommand{\instLjubljanaJSI}{J. Stefan Institute, 1000 Ljubljana, Slovenia}
\newcommand{\instKyiv}{Taras Shevchenko National Univ. of Kiev, Kiev, Ukraine}
\newcommand{\instTata}{Tata Institute of Fundamental Research, Mumbai 400005, India}
\newcommand{\instTUM}{Department of Physics, Technische Universit\"{a}t M\"{u}nchen, 85748 Garching, Germany}
\newcommand{\instECUTUM}{Excellence Cluster Universe, Technische Universit\"{a}t M\"{u}nchen, 85748 Garching, Germany}
\newcommand{\instTelAviv}{Tel Aviv University, School of Physics and Astronomy, Tel Aviv, 69978, Israel}
\newcommand{\instToho}{Toho University, Funabashi 274-8510, Japan}
\newcommand{\instTohoku}{Department of Physics, Tohoku University, Sendai 980-8578, Japan}
\newcommand{\instTitech}{Tokyo Institute of Technology, Tokyo 152-8550, Japan}
\newcommand{\instTokyoMetropolitan}{Tokyo Metropolitan University, Tokyo 192-0397, Japan}
\newcommand{\instUAS}{Universidad Autonoma de Sinaloa, Sinaloa 80000, Mexico}
\newcommand{\instNapoliUNIV}{Dipartimento di Scienze Fisiche, Universit\`{a} di Napoli Federico II, I-80126 Napoli, Italy}
\newcommand{\instNapoliUNIVA}{Dipartimento di Agraria, Universit\`{a} di Napoli Federico II, I-80055 Portici (NA), Italy}
\newcommand{\instPadovaUNIV}{Dipartimento di Fisica e Astronomia, Universit\`{a} di Padova, I-35131 Padova, Italy}
\newcommand{\instPerugiaUNIV}{Dipartimento di Fisica, Universit\`{a} di Perugia, I-06123 Perugia, Italy}
\newcommand{\instPisaUNIV}{Dipartimento di Fisica, Universit\`{a} di Pisa, I-56127 Pisa, Italy}
\newcommand{\instRomaUNIV}{Universit\`{a} di Roma ``La Sapienza,'' I-00185 Roma, Italy}
\newcommand{\instRomaTreUNIV}{Dipartimento di Matematica e Fisica, Universit\`{a} di Roma Tre, I-00146 Roma, Italy}
\newcommand{\instTorinoUNIV}{Dipartimento di Fisica, Universit\`{a} di Torino, I-10125 Torino, Italy}
\newcommand{\instTriesteUNIV}{Dipartimento di Fisica, Universit\`{a} di Trieste, I-34127 Trieste, Italy}
\newcommand{\instMontreal}{Universit\'{e} de Montr\'{e}al, Physique des Particules, Montr\'{e}al, Qu\'{e}bec, H3C 3J7, Canada}
\newcommand{\instIPHC}{Universit\'{e} de Strasbourg, CNRS, IPHC, UMR 7178, 67037 Strasbourg, France}
\newcommand{\instAdelaide}{Department of Physics, University of Adelaide, Adelaide, South Australia 5005, Australia}
\newcommand{\instBonn}{University of Bonn, 53115 Bonn, Germany}
\newcommand{\instUBC}{University of British Columbia, Vancouver, British Columbia, V6T 1Z1, Canada}
\newcommand{\instCincinnati}{University of Cincinnati, Cincinnati, Ohio 45221, U.S.A.}
\newcommand{\instFlorida}{University of Florida, Gainesville, Florida 32611, U.S.A.}
\newcommand{\instHamburg}{University of Hamburg, 20148 Hamburg, Germany}
\newcommand{\instHawaii}{University of Hawaii, Honolulu, Hawaii 96822, U.S.A.}
\newcommand{\instHeidelberg}{University of Heidelberg, 68131 Mannheim, Germany}
\newcommand{\instLjubljanaUniLJ}{Faculty of Mathematics and Physics, University of Ljubljana, 1000 Ljubljana, Slovenia}
\newcommand{\instLouisville}{University of Louisville, Louisville, Kentucky 40292, U.S.A.}
\newcommand{\instMalaya}{National Centre for Particle Physics, University Malaya, 50603 Kuala Lumpur, Malaysia}
\newcommand{\instLjubljanaUM}{University of Maribor, 2000 Maribor, Slovenia}
\newcommand{\instMelbourne}{School of Physics, University of Melbourne, Victoria 3010, Australia}
\newcommand{\instMississippi}{University of Mississippi, University, Mississippi 38677, U.S.A.}
\newcommand{\instUOM}{University of Miyazaki, Miyazaki 889-2192, Japan}
\newcommand{\instNovaGorica}{University of Nova Gorica, 5000 Nova Gorica, Slovenia}
\newcommand{\instPittsburgh}{University of Pittsburgh, Pittsburgh, Pennsylvania 15260, U.S.A.}
\newcommand{\instUSTC}{University of Science and Technology of China, Hefei 230026, China}
\newcommand{\instSAlabama}{University of South Alabama, Mobile, Alabama 36688, U.S.A.}
\newcommand{\instSCarolina}{University of South Carolina, Columbia, South Carolina 29208, U.S.A.}
\newcommand{\instSydney}{School of Physics, University of Sydney, New South Wales 2006, Australia}
\newcommand{\instTabuk}{Department of Physics, Faculty of Science, University of Tabuk, Tabuk 71451, Saudi Arabia}
\newcommand{\instUTokyo}{Department of Physics, University of Tokyo, Tokyo 113-0033, Japan}
\newcommand{\instIPMU}{Kavli Institute for the Physics and Mathematics of the Universe (WPI), University of Tokyo, Kashiwa 277-8583, Japan}
\newcommand{\instVictoria}{University of Victoria, Victoria, British Columbia, V8W 3P6, Canada}
\newcommand{\instVPI}{Virginia Polytechnic Institute and State University, Blacksburg, Virginia 24061, U.S.A.}
\newcommand{\instWayneState}{Wayne State University, Detroit, Michigan 48202, U.S.A.}
\newcommand{\instYamagata}{Yamagata University, Yamagata 990-8560, Japan}
\newcommand{\instYerevan}{Alikhanyan National Science Laboratory, Yerevan 0036, Armenia}
\newcommand{\instYonsei}{Yonsei University, Seoul 03722, South Korea}
\affiliation{\instSinica}
\affiliation{\instBeihang}
\affiliation{\instBUAP}
\affiliation{\instBNL}
\affiliation{\instBINP}
\affiliation{\instCMU}
\affiliation{\instCPPM}
\affiliation{\instCinvestavIPN}
\affiliation{\instPrague}
\affiliation{\instChiangMai}
\affiliation{\instChiba}
\affiliation{\instChonnam}
\affiliation{\instConacyt}
\affiliation{\instDESY}
\affiliation{\instDuke}
\affiliation{\instDuyTan}
\affiliation{\instENEA}
\affiliation{\instEri}
\affiliation{\instJuelich}
\affiliation{\instFuJen}
\affiliation{\instFudan}
\affiliation{\instGoettingen}
\affiliation{\instGifu}
\affiliation{\instSOKENDAI}
\affiliation{\instGyeongsang}
\affiliation{\instHanyang}
\affiliation{\instKEK}
\affiliation{\instJPARC}
\affiliation{\instHSE}
\affiliation{\instIISER}
\affiliation{\instIITBhubaneswar}
\affiliation{\instIITGuwahati}
\affiliation{\instIITHyderabad}
\affiliation{\instIITMadras}
\affiliation{\instIndiana}
\affiliation{\instIHEPRussia}
\affiliation{\instHEPHYVienna}
\affiliation{\instIHEPChina}
\affiliation{\instChennai}
\affiliation{\instIPP}
\affiliation{\instIOP}
\affiliation{\instIFIC}
\affiliation{\instFrascati}
\affiliation{\instNapoliINFN}
\affiliation{\instPadovaINFN}
\affiliation{\instPerugiaINFN}
\affiliation{\instPisaINFN}
\affiliation{\instRomaINFN}
\affiliation{\instRomaTreINFN}
\affiliation{\instTorinoINFN}
\affiliation{\instTriesteINFN}
\affiliation{\instJAEA}
\affiliation{\instMainz}
\affiliation{\instGiessen}
\affiliation{\instKarlsruhe}
\affiliation{\instKennesaw}
\affiliation{\instKitasato}
\affiliation{\instKISTI}
\affiliation{\instKorea}
\affiliation{\instKSU}
\affiliation{\instKyotoU}
\affiliation{\instKyungpook}
\affiliation{\instLAL}
\affiliation{\instLPI}
\affiliation{\instLNNU}
\affiliation{\instLMU}
\affiliation{\instLuther}
\affiliation{\instMNITJaipur}
\affiliation{\instMPP}
\affiliation{\instMPGHLL}
\affiliation{\instMcGill}
\affiliation{\instMETU}
\affiliation{\instMEPhI}
\affiliation{\instNagoya}
\affiliation{\instNagoyaKMI}
\affiliation{\instNaraWu}
\affiliation{\instUNAM}
\affiliation{\instNTUTaiwan}
\affiliation{\instNUUTaiwan}
\affiliation{\instKrakow}
\affiliation{\instNiigata}
\affiliation{\instNSU}
\affiliation{\instOkinawa}
\affiliation{\instOsakaCity}
\affiliation{\instRCNP}
\affiliation{\instPNNL}
\affiliation{\instPanjab}
\affiliation{\instPeking}
\affiliation{\instPanjabPAU}
\affiliation{\instRIKEN}
\affiliation{\instXavier}
\affiliation{\instSeoul}
\affiliation{\instShandong}
\affiliation{\instSPU}
\affiliation{\instSoochow}
\affiliation{\instSoongsil}
\affiliation{\instLjubljanaJSI}
\affiliation{\instKyiv}
\affiliation{\instTata}
\affiliation{\instTUM}
\affiliation{\instECUTUM}
\affiliation{\instTelAviv}
\affiliation{\instToho}
\affiliation{\instTohoku}
\affiliation{\instTitech}
\affiliation{\instTokyoMetropolitan}
\affiliation{\instUAS}
\affiliation{\instNapoliUNIV}
\affiliation{\instPadovaUNIV}
\affiliation{\instPerugiaUNIV}
\affiliation{\instPisaUNIV}
\affiliation{\instRomaUNIV}
\affiliation{\instRomaTreUNIV}
\affiliation{\instTorinoUNIV}
\affiliation{\instTriesteUNIV}
\affiliation{\instMontreal}
\affiliation{\instIPHC}
\affiliation{\instAdelaide}
\affiliation{\instBonn}
\affiliation{\instUBC}
\affiliation{\instCincinnati}
\affiliation{\instFlorida}
\affiliation{\instHamburg}
\affiliation{\instHawaii}
\affiliation{\instHeidelberg}
\affiliation{\instLjubljanaUniLJ}
\affiliation{\instLouisville}
\affiliation{\instMalaya}
\affiliation{\instLjubljanaUM}
\affiliation{\instMelbourne}
\affiliation{\instMississippi}
\affiliation{\instUOM}
\affiliation{\instNovaGorica}
\affiliation{\instPittsburgh}
\affiliation{\instUSTC}
\affiliation{\instSAlabama}
\affiliation{\instSCarolina}
\affiliation{\instSydney}
\affiliation{\instTabuk}
\affiliation{\instUTokyo}
\affiliation{\instIPMU}
\affiliation{\instVictoria}
\affiliation{\instVPI}
\affiliation{\instWayneState}
\affiliation{\instYamagata}
\affiliation{\instYerevan}
\affiliation{\instYonsei}
  \author{F.~Abudin{\'e}n}\affiliation{\instTriesteINFN} 
  \author{I.~Adachi}\affiliation{\instKEK}\affiliation{\instSOKENDAI} 
  \author{R.~Adak}\affiliation{\instFudan} 
  \author{K.~Adamczyk}\affiliation{\instKrakow} 
  \author{P.~Ahlburg}\affiliation{\instBonn} 
  \author{J.~K.~Ahn}\affiliation{\instKorea} 
  \author{H.~Aihara}\affiliation{\instUTokyo} 
  \author{N.~Akopov}\affiliation{\instYerevan} 
  \author{A.~Aloisio}\affiliation{\instNapoliUNIV}\affiliation{\instNapoliINFN} 
  \author{F.~Ameli}\affiliation{\instRomaINFN} 
% ...
\collaboration{Belle II Collaboration}

%% file: acknowledgements.tex
We thank the SuperKEKB group for the excellent operation of the
accelerator; the KEK cryogenics group for the efficient
operation of the solenoid; and the KEK computer group for
on-site computing support.
This work was supported by the following funding sources:
Science Committee of the Republic of Armenia Grant No. 18T-1C180;
Australian Research Council and research grant Nos.
DP180102629, 
DP170102389, 
DP170102204, 
DP150103061, 
FT130100303, 
and
FT130100018; 
Austrian Federal Ministry of Education, Science and Research, and
Austrian Science Fund No. P 31361-N36; 
Natural Sciences and Engineering Research Council of Canada, Compute Canada and CANARIE;
Chinese Academy of Sciences and research grant No. QYZDJ-SSW-SLH011,
National Natural Science Foundation of China and research grant Nos.
11521505,
11575017,
11675166,
11761141009,
11705209,
and
11975076,
LiaoNing Revitalization Talents Program under contract No. XLYC1807135,
Shanghai Municipal Science and Technology Committee under contract No. 19ZR1403000,
Shanghai Pujiang Program under Grant No. 18PJ1401000,
and the CAS Center for Excellence in Particle Physics (CCEPP);
the Ministry of Education, Youth and Sports of the Czech Republic under Contract No.~LTT17020 and 
Charles University grants SVV 260448 and GAUK 404316;
European Research Council, 7th Framework PIEF-GA-2013-622527, 
Horizon 2020 Marie Sklodowska-Curie grant agreement No. 700525 `NIOBE,' 
and
Horizon 2020 Marie Sklodowska-Curie RISE project JENNIFER grant agreement No. 644294 (European grants);
L'Institut National de Physique Nucl\'{e}aire et de Physique des Particules (IN2P3) du CNRS (France);
BMBF, DFG, HGF, MPG and AvH Foundation (Germany);
Department of Atomic Energy and Department of Science and Technology (India);
Israel Science Foundation grant No. 2476/17
and
United States-Israel Binational Science Foundation grant No. 2016113;
Istituto Nazionale di Fisica Nucleare and the research grants BELLE2;
Japan Society for the Promotion of Science,  Grant-in-Aid for Scientific Research grant Nos.
16H03968, 
16H03993, 
16H06492,
16K05323, 
17H01133, 
17H05405, 
18K03621, 
18H03710, 
18H05226,
19H00682, 
26220706,
and
26400255,
the National Institute of Informatics, and Science Information NETwork 5 (SINET5), 
and
the Ministry of Education, Culture, Sports, Science, and Technology (MEXT) of Japan;  
National Research Foundation (NRF) of Korea Grant Nos.
2016R1\-D1A1B\-01010135,
2016R1\-D1A1B\-02012900,
2018R1\-A2B\-3003643,
2018R1\-A6A1A\-06024970,
2018R1\-D1A1B\-07047294,
2019K1\-A3A7A\-09033840,
and
2019R1\-I1A3A\-01058933,
Radiation Science Research Institute,
Foreign Large-size Research Facility Application Supporting project,
the Global Science Experimental Data Hub Center of the Korea Institute of Science and Technology Information
and
KREONET/GLORIAD;
Universiti Malaya RU grant, Akademi Sains Malaysia and Ministry of Education Malaysia;
Frontiers of Science Program contracts
FOINS-296,
CB-221329,
CB-236394,
CB-254409,
and
CB-180023, and the Thematic Networks program (Mexico);
the Polish Ministry of Science and Higher Education and the National Science Center;
the Ministry of Science and Higher Education of the Russian Federation,
Agreement 14.W03.31.0026;
University of Tabuk research grants
S-1440-0321, S-0256-1438, and S-0280-1439 (Saudi Arabia);
Slovenian Research Agency and research grant Nos.
J1-9124
and
P1-0135; 
Agencia Estatal de Investigacion, Spain grant Nos.
FPA2014-55613-P
and
FPA2017-84445-P,
and
CIDEGENT/2018/020 of Generalitat Valenciana;
Ministry of Science and Technology and research grant Nos.
MOST106-2112-M-002-005-MY3
and
MOST107-2119-M-002-035-MY3, 
and the Ministry of Education (Taiwan);
Thailand Center of Excellence in Physics;
TUBITAK ULAKBIM (Turkey);
Ministry of Education and Science of Ukraine;
the US National Science Foundation and research grant Nos.
PHY-1807007 
and
PHY-1913789, 
and the US Department of Energy and research grant Nos.
DE-AC06-76RLO1830, 
DE-SC0007983, 
DE-SC0009824, 
DE-SC0009973, 
DE-SC0010073, 
DE-SC0010118, 
DE-SC0010504, 
DE-SC0011784, 
DE-SC0012704; 
and
the National Foundation for Science and Technology Development (NAFOSTED) 
of Vietnam under contract No 103.99-2018.45.